\documentclass[fleqn,12pt,twoside]{article}
\usepackage[headings]{espcrc1}

\usepackage{graphicx}
\usepackage[figuresright]{rotating}

\newcommand{\AmS}{{\protect\the\textfont2
  A\kern-.1667em\lower.5ex\hbox{M}\kern-.125emS}}

\title{Anisotropic Flow in $\sqrt{s_{NN}}$ = 200 GeV 
  Cu+Cu and Au+Au collisions at PHENIX}

\author{
  H. Masui \address[TSUKUBA]{Inst. of Physics, Univ. of Tsukuba, 
  Tenno-dai 1-1-1, Tsukuba, Ibaraki, Japan} (for the PHENIX 
  \thanks{ For the full list of PHENIX authors and acknowledgements, 
    see Appendix 'Collaborations' of this volume.}
      collaboration)
}

\runtitle{Anisotropic Flow in $\sqrt{s_{NN}}$ = 200 GeV Cu+Cu and Au+Au collisions at PHENIX}

\runauthor{H. Masui, for the PHENIX collaboration}

\begin{document}

\maketitle

\begin{abstract}
  We report the measurement of anisotropic flow at RHIC - PHENIX experiment.
  We present the $v_4$ results at $\sqrt{s_{NN}}$ = 200 GeV in Au+Au collision.
  The scaling ratio of $v_4/(v_2)^2$ is about 1.5 and it is found to be smaller than the 
  prediction from simple coalescence model.
  The $v_2$ for high $p_T$ identified particles ($\sim$ 5 GeV/c) measured with Aerogel Cherenkov Counter 
  are presented. We discuss the constituent quark scaling of $v_2$ for identified particles. 
  We also report the first observation of $v_2$ for inclusive charged hadrons as well as identified 
  hadrons at $\sqrt{s_{NN}}$ = 200 GeV in Cu+Cu collisions. The system size dependence of $v_2$ and 
  scaling properties are discussed.
\end{abstract}

\section{Introduction}
 Anisotropic Flow, an anisotropy of the azimuthal distribution of emitted particles in 
 momentum space with respect to the reaction plane, is a sensitive tool to study the 
 early stage of high energy heavy ion collisions at RHIC. It is commonly studied by 
 measuring the Fourier harmonics ($v_n$) of this distribution \cite{Art01}. 
 Elliptic Flow ($v_2$) in Au+Au collisions is well studied at RHIC and is thought to reflect 
 conditions from the early stage of the collision.

 It has been reported that anisotropic flow can be reasonably described by the hydrodynamical 
 model up to 2 GeV/c \cite{Houv01}. Hydrodynamics also predict several scaling relations, 
 such as the scaling of $v_4/(v_2)^2$, eccentricity scaling and the scaling with system size.
 It is crucial to test these scaling relations in order to investigate the properties of the hot and 
 dense matter created at an early stage of the heavy ion collisions.

\section{Analysis}

The data taken by the PHENIX central arm for Au+Au and Cu+Cu collisions at $\sqrt{s_{NN}}$ = 
200 GeV were used in this analysis. The PHENIX central arm covered half of azimuth,
pseudorapidity ($\eta$) from -0.35 to 0.35 and the low transverse momentum ($p_T$) cut off was 
0.2 GeV/c. Identified hadrons are measured by the high resolution Time-of-Flight (TOF) and 
the threshold type Aerogel Cherenkov Counter.

We used the event plane method \cite{Art01} for anisotropic flow measurement. Event plane was 
determined at Beam-Beam Counter (BBC) \cite{PHENIX01} which covered full azimuth and 
$|\eta|$ from 3.0 to 3.9. The large rapidity gap between the PHENIX central arm and BBC 
($\Delta\eta \sim$ 3) reduces the non-flow contribution to the two-particle azimuthal 
correlations.

\section{$v_4\{EP_2\}$}
\begin{figure}[htbp]
    \includegraphics[scale=0.7,clip]{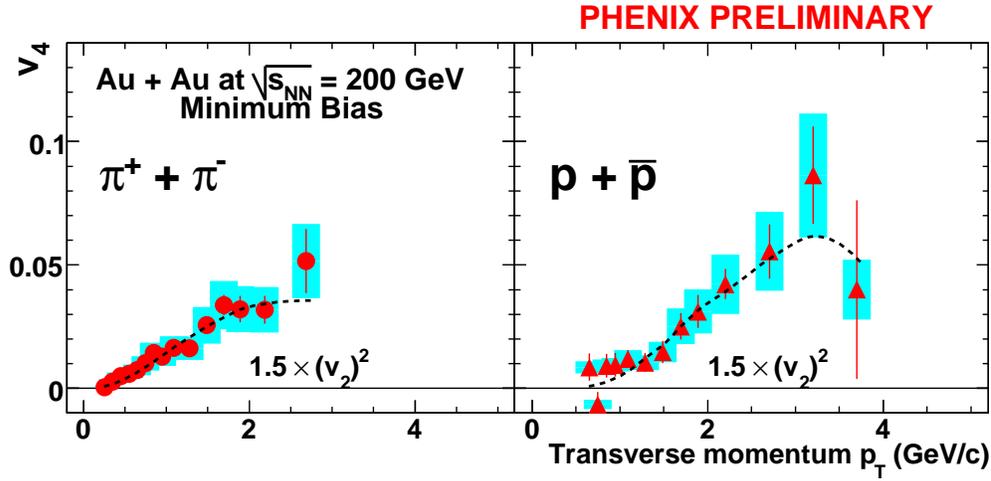}
    \vskip -10mm
    \caption{
      The minimum bias $v_4$ for identified pion (left) and proton (right) in Au+Au collision with respect 
        to the second harmonic event plane as a function of $p_T$. The error bars are statistical error, and 
        shaded boxes are systematic error.  The dashed curves are 1.5 $\times$ $(v_2)^2$.
    }
    \label{fig:fig1}
\end{figure}

Figure \ref{fig:fig1} shows the $v_4$ for charged pions and protons with respect to 
the second harmonic BBC event plane in minimum bias event (0 - 93 \% centrality).
The $p_T$ dependence of $v_4$ is very similar to the $v_2$, that is, proton $v_4$ is smaller than 
pions in low $p_T$ and in higher $p_T$ the magnitude of proton $v_4$ is same as or larger than pions. 
The ratio, $v_4/(v_2)^2$, has been found to be scaled with the proportionality constant 1.2 by STAR collaboration \cite{STAR01}.
We observe the ratio $v_4/(v_2)^2$ is scaled by 1.5 but this result is consistent with that of \cite{STAR01} 
within systematic uncertainties. A simple parton coalescence model predicts the ratio $v_4/(v_2)^2$ for both 
mesons and baryons as \cite{Kolb01}:
\begin{equation}
  v_4^m/(v_2^m)^2 \approx 1/4 + 1/2 (v_4^q/(v_2^q)^2), \hspace{2cm}
  v_4^b/(v_2^b)^2 \approx 1/3 + 1/3 (v_4^q/(v_2^q)^2)
\end{equation}
where $v_2^m$, $v_2^b$ and $v_2^q$ are meson, baryon and parton $v_2$, respectively. If one assume that 
the following scaling relation \cite{AMPT01} for higher order parton anisotropic flow, $v_4^q \sim (v_2^q)^2$, then 
one get $1/4 + 1/2 = 3/4$ for mesons, $ 1/3 + 1/3 = 2/3$ for baryons. These predictions underestimate the experimental 
results, and this would indicate the $v_4^q$ is larger than simple parton $(v_2^q)^2$. Recently, ideal-fluid 
model suggests that $v_4/(v_2)^2$ = 0.5 and this should become more and more accurate for higher $p_T$ 
\cite{Hydro01}. Further study of higher harmonics at RHIC would be very interesting to understand properties 
of the system.

\section{Identified hadron $v_2$}
\begin{figure}[htbp]
\begin{minipage}[t]{100mm}
  \begin{center}
  \includegraphics[bb = 0 0 525 377,width=100mm,clip]{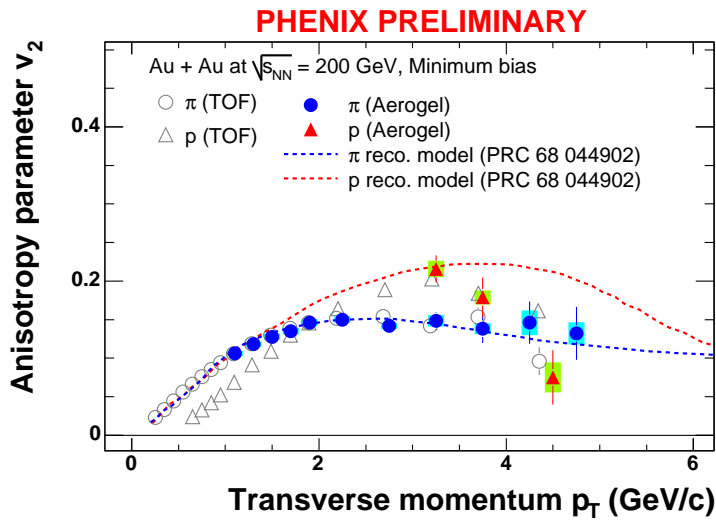}
  \end{center}
\end{minipage}
\hspace{2mm}
\begin{minipage}[t]{58mm}
\vskip -70mm
  \caption{
    $v_2$($p_T$) for identified pion (circles) and proton (triangles) measured by Aerogel Cherenkov 
      Counter. $v_2$ measured by Time-of-Flight are also plotted by open data points for comparison.
      The dashed curves are the calculation from recombination model \cite{RECO01}.
  }
  \label{fig:fig2}
\end{minipage}
\end{figure}

Figure \ref{fig:fig2} shows $v_2$ for identified pion and proton measured by Aerogel as a function of 
$p_T$.
We extended the $p_T$ reach by Aerogel counter up to 5 GeV/c for identified pion and proton. Pion $v_2$ 
is quite consistent with the prediction of recombination model \cite{RECO01} as shown by dashed curves in figure 
\ref{fig:fig2}.  The proton $v_2$, however, seems to decrease for $p_T$ = 4 - 5 GeV/c and clearly deviates from 
recombination model. The mass splitting of $v_2$ at low $p_T$ ($p_T <$  2 GeV/c) are also not described by the 
recombination model.  The similar trend can also see in TOF data.  We still need further study in this $p_T$ and 
higher with Aerogel to investigate the production mechanism of protons.

\section{System size dependence of $v_2$: Cu+Cu vs Au+Au}
\begin{figure}[htbp]
\begin{minipage}[t]{78mm}
  \begin{center}
  \includegraphics[width=78mm]{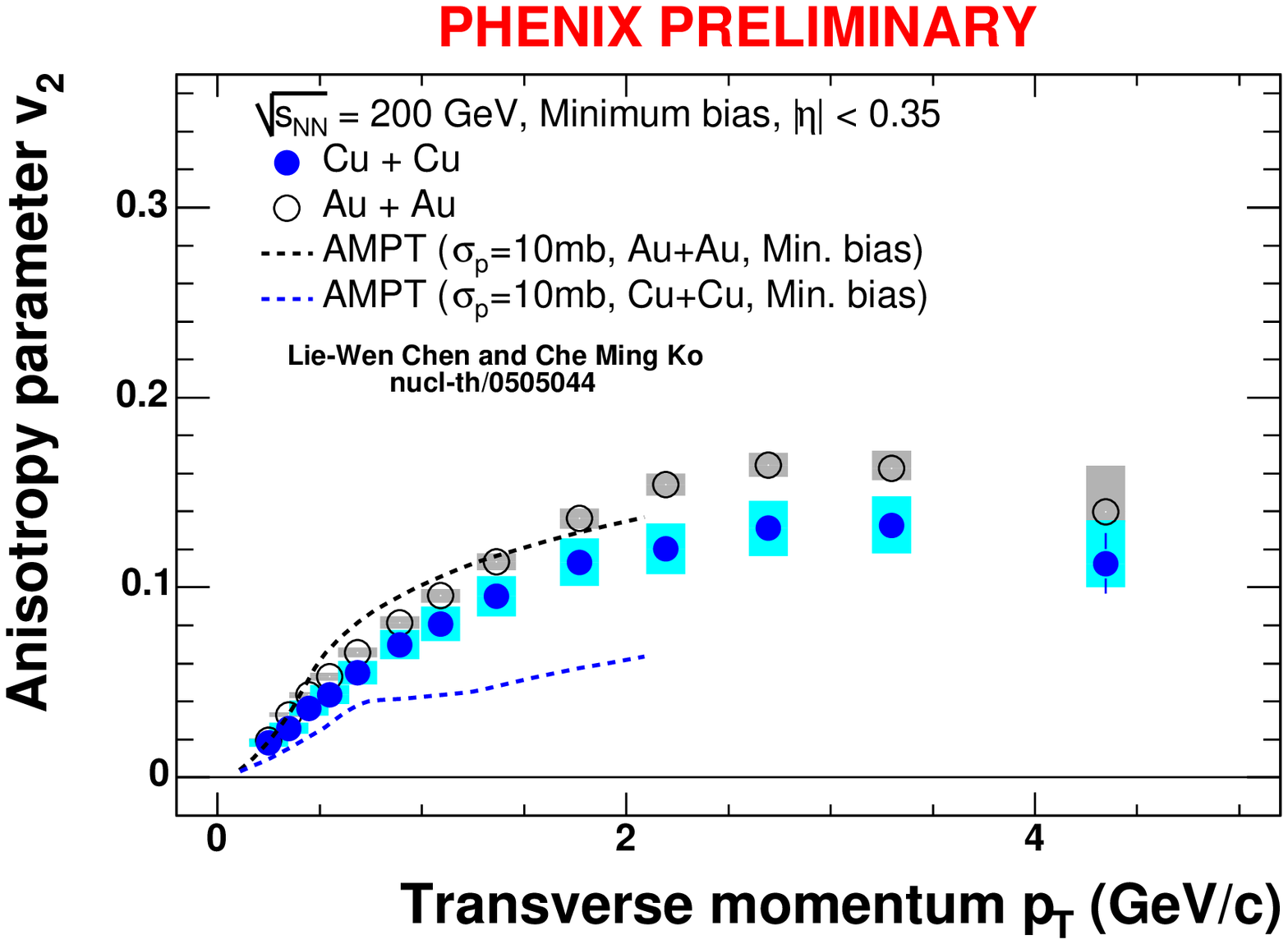}
  \end{center}
  \vskip -10mm
\caption{
  $v_2$ for inclusive charged hadrons in Cu+Cu and Au+Au collisions for minimum bias 
    event as a function of $p_T$. The dashed curves are predictions by AMPT model \cite{AMPT02}.
}
\label{fig:fig3}
\end{minipage}
\hspace{2mm}
\begin{minipage}[htbp]{78mm}
  \vskip -18mm
  \begin{center}
  \includegraphics[bb = 0 0 531 355,width=78mm,clip]{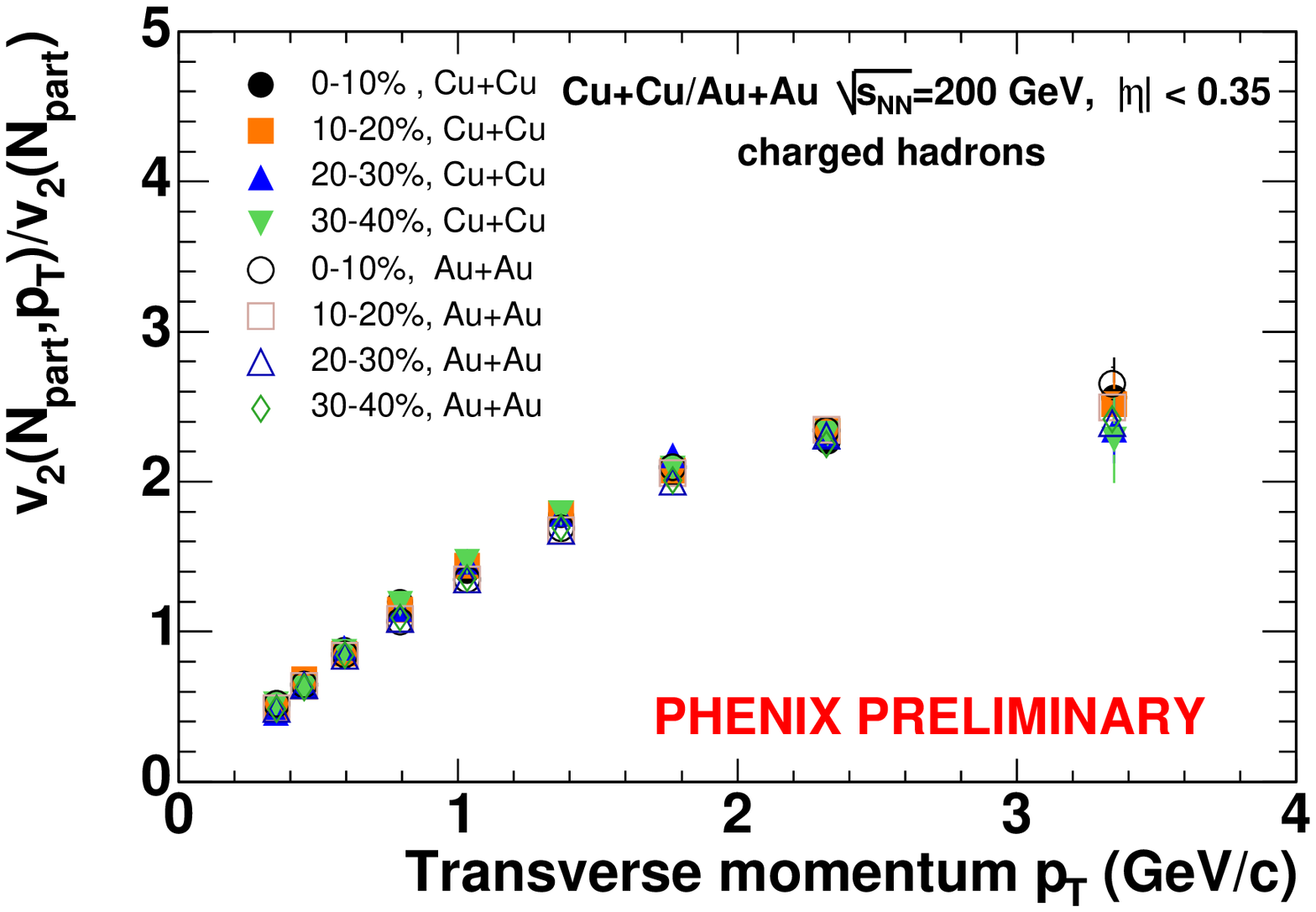}
  \end{center}
  \vskip -12mm
\caption{
  Eccentricity scaling (the ratio of differential $v_2$($p_T$) to integrated $v_2$($p_T$)) for 
    inclusive charged hadrons in Cu+Cu and Au+Au collisions for 0-10 \%, 10-20 \%, 20-30 \% and 30-40 \%.
}
\label{fig:fig4}
\end{minipage}
\end{figure}
The inclusive charged hadron $v_2$($p_T$) in Cu+Cu collisions compared to that in Au+Au are shown in figure \ref{fig:fig3}.
The $p_T$ dependence of $v_2$ in Cu+Cu is similar to that in Au+Au and the magnitude of $v_2$ is smaller than in 
Au+Au about 15 \% in minimum bias event. One of the parton cascade model, AMPT, predicts that the $v_2$ in Cu+Cu 
is scaled with the system size linearly so that $v_2$(Cu)$\sim$ 0.3 $\cdot$ $v_2$(Au) \cite{AMPT02}. 
However, the data seems to not scale with the system size linearly.  
The difference between data and model prediction might be due to non-flow contribution but it is necessary to have more 
study to understand the system size dependence of $v_2$.

Figure \ref{fig:fig4} shows the scaling of $v_2$, which is the ratio of differential $v_2$ to integrated one, for charged hadrons 
in Cu+Cu and Au+Au collisions as a function of $p_T$. We use the fact that integrated $v_2$ ($v_2(N_{part})$) approximately scales 
with eccentricity ($\epsilon$).  Thus the scaling of $v_2$ is approximately equal to $v_2/\epsilon$. 
This scaling has advantages which could reduce the systematic errors related to eccentricity calculation and 
event plane resolution. One can see that the scaling of $v_2$ is roughly independent of the system size from central 
to mid-central collisions. This would suggest that hydrodynamical regime is reached in central to mid-central Cu+Cu 
collisions.

\section{Conclusions}

We measure the anisotropic flow for various particle species in Au+Au and Cu+Cu collisions 
at $\sqrt{s_{NN}}$ = 200 GeV at RHIC-PHENIX experiment. The scaling ratio $v_4/(v_2)^2$ for 
both unidentified and identified hadrons are 1.5 and it is consistent with the previous measurement 
by STAR experiment. Proton $v_2$ seems to decrease and deviate from recombination model 
for $p_T$ = 4 - 5 GeV/c, while pion $v_2$ has good agreement with recombination model.
The $v_2$ in Cu+Cu is smaller than that in Au+Au.
The eccentricity scaling of $v_2$ for Cu+Cu and Au+Au is roughly independent of the system size and this suggest that 
hydro regime is reached in central to mid-central Cu+Cu collisions at RHIC.

\end{document}